# Control of Domain Wall Position by Electrical Current in Structured Co/Ni Wire with Perpendicular Magnetic Anisotropy

Tomohiro Koyama, Gen Yamada, Hironobu Tanigawa, Shinya Kasai, Norikazu Ohshima[1], Shunsuke Fukami[1], Nobuyuki Ishiwata[1], Yoshinobu Nakatani[2] and Teruo Ono

*Institute for Chemical Research, Kyoto University, Uji 611-0011, Japan*

[1]*Device Platforms Research Laboratories, NEC Corporation,*

*1120 Shimokuzawa, Sagamihara, 229-1198, Japan*

[2]*University of Electro-communications, Chofu, 182-8585, Tokyo, Japan*

**Abstract**

We report the direct observation of the current-driven domain wall (DW) motion by magnetic force microscopy in a structured Co/Ni wire with perpendicular magnetic anisotropy. The wire has notches to define the DW position. It is demonstrated that single current pulses can precisely control the DW position from notch to notch with high DW velocity of 40 m/s.

Domain wall (DW) motion in magnetic wires by current injection, first predicted by Berger [1], is now investigated extensively from both experimental [2–22] and theoretical [23–29] points of view, since in addition to exciting fundamental physics, novel applications based on the current-driven DW motion have been proposed [30-32]. These novel spintronic devices require a low drive current for DW motion and a high stability of DW position to store the information. However, these two conditions are not fulfilled together for NiFe wires, which have been investigated most intensively. A high current density of $1 \times 10^{12}$ A/m$^2$ is needed even for a small DW pinning field of 5 Oe, and the threshold current density increases with the pinning field [22]. Recently, it was shown by the micromagnetics simulations that the wire with perpendicular magnetic anisotropy is a good candidate which fulfills the above two conditions [33, 34]. We have shown that, in a CoCrPt wire with perpendicular magnetic anisotropy, a DW was successfully displaced by an electric current in spite of the large pinning field of 500 Oe [35]. However, the DW velocity was only 0.05 m/s in spite of the high threshold current density of $1.3 \times 10^{12}$ A/m$^2$, and the DW motion seems to be influenced by the defects in the wire.

In this paper, we demonstrate the fast and precise control of the DW position by current pulses in a Co/Ni wire with notches. The depinning field from the notch in the wire is about 400 Oe for the sample reported in this letter. Magnetic force microscopy (MFM) observation shows that the position of a single DW can be precisely controlled from notch to notch by the injection of a current pulse in spite of the large depinning field.

A schematic illustration of the sample with a scanning electron microscope image is shown in Fig. 1. The Co/Ni wire has four notches to define the DW position.

The widths of the widest and the narrowest part of the wire are 94 nm and 72 nm, respectively. The distance between notches is 400 nm. The wire was patterned by electron beam lithography from [Co(0.3 nm)/Ni(0.9 nm)]$_4$/Co(0.3 nm) films deposited on silicon substrates by alternative dc magnetron sputtering of Co and Ni layers. The film had the saturation magnetization of 680 emu/cm$^3$ with the perpendicular magnetic anisotropy of $3.8 \times 10^6$ erg/cm$^3$.

MFM observation of the current-driven DW motion in the Co/Ni wire was performed in the following way. First, a magnetic field of 2 kOe perpendicular to the substrate plane was applied in order to align the magnetization in one direction. Then, a current pulse with the duration of 50 ns was applied though the Au/Cr wire from the electrode *A* to *B* to nucleate a domain wall by the Oersted field, value of which was estimated to be 630 Oe. Then, the current-driven DW experiments were performed by injecting a current pulse through the Co/Ni wire by using the electrodes *A* and *C*. The MFM observations were carried out in the absence of a magnetic field.

Figure 2(a) shows an MFM image after the DW introduction. The part of the Co/Ni wire is surrounded by the white dashed lines, and the region of the Au/Cr wire is shaded. The dark contrast in the left part of the second notch corresponds to the downward magnetization, while the bright contrast in the right part corresponds to the upward magnetization, indicating that a DW is trapped at the second notch. Figures 2(b) - 2(f) are a sequence of MFM images recorded after the injection of single current pulses, $1.4 \times 10^{12}$ A/m$^2$ and 10 ns long, with alternative current polarity. It is found that the current pulse can displace the DW back and forth between the second and the third notches.

We examined the current-driven DW motion between the second and the third notches for various combinations of the current density $J$ and the pulse duration. The experiment was performed 40 times for each combination. It was confirmed that the DW always moved opposite to the current direction whenever it changed its position by the injection of a current pulse. The results are summarized in Fig. 3. Here, we define the success probability as the ratio of the number of the successful DW motion to the total trial number of 40 times. The successful DW motion means the DW displacement from the second (third) to the third (second) notch by the current injection. The threshold current density can be determined to be $1.3 \times 10^{12}$ A/m$^2$, at which current density the success probabilities show the abrupt increases. The low success probability for the pulse duration of 5 ns indicates that the duration is not long enough to displace the DW from notch to notch. The reason why the success probability for the 20 ns-duration is lower than that for the 10 ns-duration above the threshold current density is that the DW sometimes moved over two notches in the case of the 20 ns-duration ( 4 times in 40 trials for $1.3 \times 10^{12}$ A/m$^2$, and 6 times in 40 trials for $1.4 \times 10^{12}$ A/m$^2$). Thus, for the current density just above the threshold, the 10 ns pulse is appropriate for the distance of 400 nm between notches, leading to the rough estimation of the average DW velocity of 40 m/s.

We have shown that a single current pulse can control precisely the DW position from notch to notch in a Co/Ni wire with perpendicular magnetic anisotropy. The large built-in DW pinning potential produced by the notch and the high DW velocity of 40 m/s certify that the Co/Ni system is a promising candidate for the domain-wall-based spintronics devices.

The present work was partly supported by NEDO Spintronics nonvolatile devices project. HT acknowledges support from JSPS Research Fellowship for Young Scientists.

**Figure captions**

Figure 1

Schematic illustration of the sample with a scanning electron microscope image.

Figure 2

MFM observations of the current-driven DW motion by injecting single current pulses, $1.4 \times 10^{12}$ A/m$^2$ and 10 ns long.

Figure 3

Success probability as a function of current density for different pulse durations (5, 10, and 20 ns).

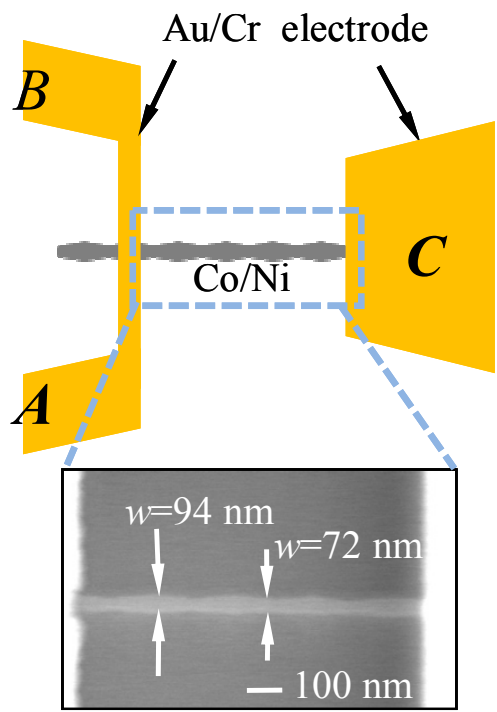

Figure 1 T. Koyama

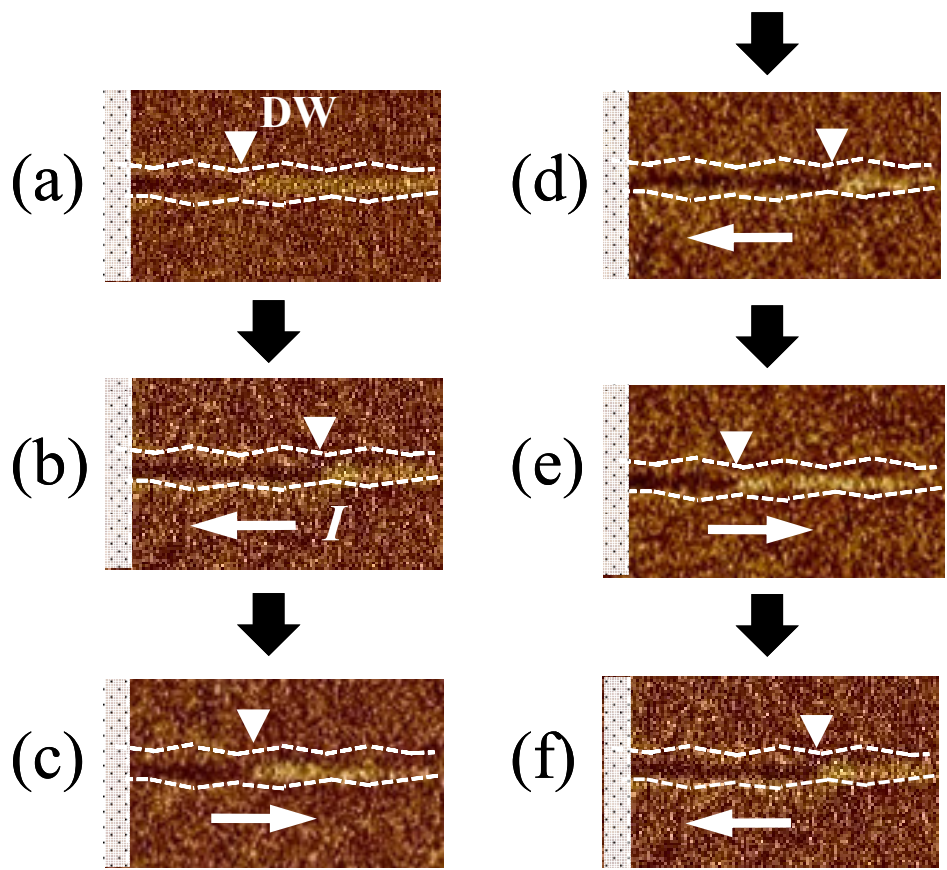

Figure 2 T. Koyama

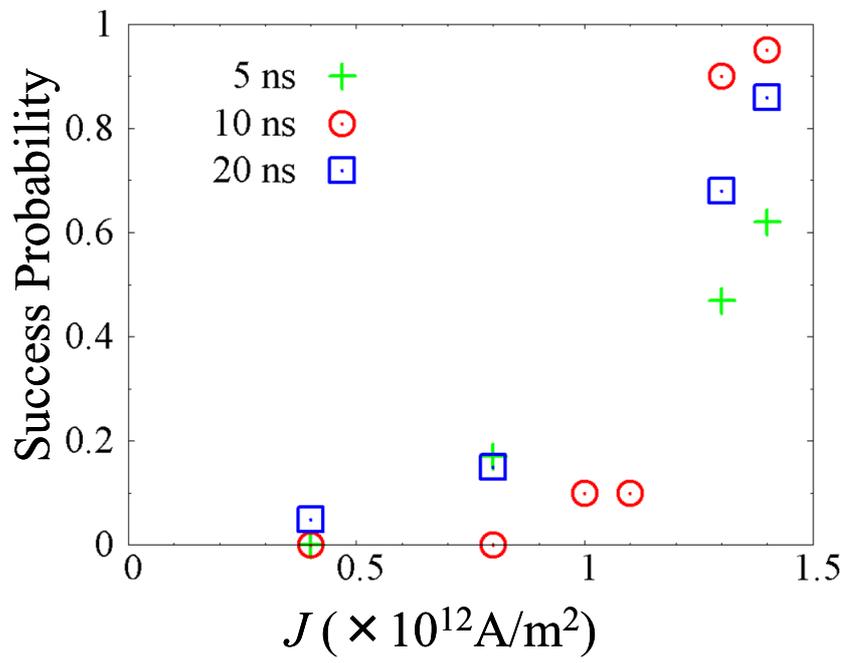

Fig.3 T. Koyama